\documentclass[twocolumn,showpacs,amsmath,amssymb,prb,%
               superscriptaddress]{revtex4}
\usepackage[T1]{fontenc}
\usepackage{times}
\usepackage{graphicx,color}
\usepackage{wasysym}
\usepackage{bm}

\begin{document}

\title{Levitation of the quantum Hall extended states in the
       $\bm{B\to}$ 0 limit} 
\author{Th.\@ Koschny}
\affiliation{Physikalisch-Technische Bundesanstalt, Bundesallee 100, 38116
Braunschweig, Germany}
\affiliation{Ames Laboratory and Dept.~Phys.~and Astronomy,
             Iowa State University, Ames, Iowa 50011, \\
             and Foundation for Research and Technology Hellas,
             71110 Heraklion, Crete, Greece}
\author{L. Schweitzer}
\affiliation{Physikalisch-Technische Bundesanstalt, Bundesallee 100, 38116
Braunschweig, Germany}

\date{\today}

\begin{abstract}
We investigate the fate of the quantum Hall extended states within a 
continuum model with spatially correlated disorder potentials. The
model can be projected onto a couple of the lowest Landau bands. Levitation
of the $n=0$ critical states is observed if at least the two
lowest Landau bands are considered.
The dependence on the magnetic length $l_B=(\hbar/(eB))^{1/2}$ and on the
correlation length of the disorder potential $\eta$ is combined into 
a single dimensionless parameter $\hat\eta=\eta/l_B$. This enables us 
to study the behavior of the critical states for vanishing magnetic field.
In the two Landau band limit, we find a disorder dependent saturation of 
the critical states' levitation which is in contrast to earlier 
propositions, but in accord with some experiments.
\end{abstract}


\maketitle


\section{Introduction}
What happens to the current carrying electronic states in a quantum Hall 
sample when the magnetic field is turned off? According to the 
prevailing view, at $B=0$ the single particle states of a 
disordered two-dimensional system are localized and no current can 
flow at zero temperatures. \cite{AALR79} A scenario, how this transition 
from the quantum Hall liquid into the insulator may take place, 
was suggested by Khmelnitskii \cite{Khm84} and Laughlin \cite{Lau84}
20 years ago. They both proposed that, in the limit $\omega_c
=eB/m\to 0$, the extended states of the n-th Landau level float up in 
energy, $E_n=\hbar\omega_c\, (n+1/2)\, (1+(\omega_c\tau)^{-2})$,
eventually crossing the Fermi level one by one, until the lowest
current carrying state gets depopulated. Here, the classical
elastic collision time $\tau$ is a function of the disorder 
strength.

Over the years, many experiments have been carried out to check 
this prediction, but produced only conflicting results,
\cite{STC95,Sea95b,Sea97a,LCSL98,DJS98,HaN99,Hea00} 
and a convergent picture seems to become apparent only recently 
(see, e.g., Yasin \textit{et al.}\cite{Yea02} and references therein). 
In essence, the levitation of the current carrying state's energies has 
been observed in various samples for very low magnetic fields. However, 
some find a saturation of the Landau level shift as $B\to 0$, 
for instance in high quality p-GaAs samples, \cite{DJS98,HaN99}
and even direct transitions to the Hall insulator from higher ($\nu>2$) 
quantum Hall plateaus have been reported.\cite{Sea95b,LCSL98,Hea00}
Therefore, the ultimate fate of $E_0$ in the limit 
$B\to 0$ remains an important and still open question. 
The answer to this long lasting problem has become even more pressing
through the observation of the apparent 'metallic' behavior 
found at $B=0$ in experiments on dilute two-dimensional
electron and hole systems.\cite{KKFPD94,PFW97,CWFZ97,HaN99} 

Despite quite some effort made in the past to provide a microscopic 
theory,\cite{And84,SR95,HY97,Fog98}  
our understanding of the levitation scenario is still limited and
mainly comes from numerical studies 
\cite{LXN96,YB96,SW97,SW98,YB99,KPS01,PS02b}
that are, however, afflicted by artifacts originating from the applied 
lattice model, and the difficulty of performing a proper limit $B\to 0$. 
The aim of our present work is to overcome these shortcomings and to study
the levitation of the current carrying states within a continuum model
that allows to take the essential limit $B\to 0$. Within our model,
which in a first step is projected onto the lowest Landau levels (LLL), 
we do not find an unbounded floating up in energy,
but a saturation of the quantum Hall extended states' levitation that 
depends on the disorder strength. The main advantages of our method
are the straight forward extension to cases where additional Landau
levels are taken into account as well as the possible identification
of those matrix elements responsible for the levitation because they 
are known analytically. 

\section{Model and Method}
We consider non-interacting electrons moving in the $x$-$y$-plane of  
a two-dimensional continuum model with a perpendicular magnetic field 
and quenched disorder.
We choose spatially correlated disorder potentials to be generated
by a random weighted sum of Gaussian shaped potential hills at
random sites distributed uniformly over the area of the sample.
This disorder potential can be considered the continuum generalization of
the correlated disorder potentials on the lattice used in our previous
work \cite{KS03}
\begin{equation}
V({\bf r})=
\frac{2W}{\eta\,\sqrt{2\pi c}}\ \sum_i\ 
\exp(-|\,{\bf r}-{\bf r}_i\,|^2/\eta^2)
\,\varepsilon_i.
\label{potential_1}
\end{equation}
The summation index $i$ runs over all scatterers at random positions 
${\bf r}_i$, the number of which is $N_s$.
The random variables ${\bf r}_i$ 
and $\varepsilon_i$ are both uncorrelated and uniformly distributed over 
the area of the system, and over the real interval [$-$1,1], respectively.
$W$ is the strength of the disorder potential and $c$ the
concentration of the scatterers.
The disorder potential is locally uniformly distributed, 
spatially Gaussian correlated,
$\mathrm{<}V({\bf r})V({\bf r}')\mathrm{>} = (W^2/3)\, %
 \exp[-|{\bf r}-{\bf r}'|^2/(2\eta^2)]$ 
with a correlation length 
$\sqrt{2}\,\eta$, has a vanishing mean 
$\mathrm{<}V({\bf r})\mathrm{>}=0$, and a uniform second moment
$\mathrm{<}V({\bf r})^2\mathrm{>}=W^2/3$.
Thus, the square of the disorder strength $W^2$ is effectively the second 
moment of the disorder potential.

Now we study the Hamiltonian 
$H_0=\frac{1}{2m}\big({\bf p}-e{\bf A}({\bf r})\big)^2$ 
of free two-dimensional electrons moving in a homogeneous perpendicular 
magnetic field $B$ given by a vector potential 
${\bf A}({\bf r})=-|B|\ y\,{\bf e}_x$ with ${\bf r}=(x,y)$
in the Landau gauge. 
For a system infinite in $y$-direction and periodic boundary conditions
in $x$-direction, $H_0$ possesses the well-known eigen base
\begin{eqnarray}
 \lefteqn{\Psi_{nk}(x,y) \ = \ }
 \label{spaghetti_base} \\
& &
 (L_x^2l_B^2\pi)^{-\frac{1}{4}}\;
 (2^n n!)^{-\frac{1}{2}}\
 H_n\!\left(\frac{y-\frac{\hbar k}{m \omega_c}}{2l_B^2}\right)\,
 e^{ikx}
 \,
 \nonumber
\end{eqnarray}
in terms of the Hermite polynomials $H_n$.
Here, $L_x$ is the finite system width in the periodic direction, 
$l_B^2=\hbar/(eB)$ the magnetic length. The non-negative integer $n$ is 
the Landau level index, and $k=2\pi\,m/L_x$ with integer $m$ represents the 
momentum in $x$-direction. 
For the following numerical evaluation we have to restrict
the momenta $k$ to a finite number, $k \in[0,k_{\rm max}]$, 
which effectively restricts the system to a finite area 
$F\approx  k_{\rm max} L_x l_B^2$
with open boundary conditions in the $y$-direction.
We rewrite the Hamiltonian 
$H=\frac{1}{2m}\big({\bf p}-e{\bf A}({\bf r})\big)^2+V({\bf r})$ for the 
disordered system in the representation of the free Hamiltonian's 
eigen base [Eq.~(\ref{spaghetti_base})],
\begin{eqnarray}
 \lefteqn{H \ =\ %
   \sum\limits_{n',\,k';\,n,\,k} %
   \big|\,n'k'\big>\, H_{nk}^{n'k'} \big<nk\,\big| %
   \quad \mbox{with} }
 \label{hamiltonian_projected_1} \\
 && H_{nk}^{n'k'}  =\ 
 \hbar\omega_c\left(n+\frac{1}{2}\right)\,\delta_{n}^{n'}\delta_{k}^{k'}
 \ +\ 
 \big<n'k'\,\big|V({\bf r})\big|\,nk\big>
 \,.
 \nonumber
\end{eqnarray}
Now the projection to a finite set of Landau levels is easily achieved by 
restricting the base and the $n$ and $n'$ sums in the Hamiltonian 
[Eq.~(\ref{hamiltonian_projected_1})] to these Landau levels. 
Because the disorder potential is a random weighted sum over Gaussians, 
we can easily compute the matrix elements of the disorder potential in the 
approximation that the area where each single Gaussian is non-zero is small
compared to the area of the sample. Then we can replace the spatial
integrations over the finite sample by infinite Gaussian integrals.

Further, we introduce dimensionless quantities by measuring all energies in
units of $\hbar\omega_c$ and all lengths in units of the magnetic length 
$l_B$. We will denote those quantities by attaching a hat to their symbols.
This way the Hamiltonian only depends on the dimensionless correlation length
parameter $\hat\eta$, the magnetic field only enters the energy and
length scales. 
It takes the form
\begin{equation*}
 \hat{H}_{nk}^{n'k'} =\,
 \left(n+\frac{1}{2}\right)\,\delta_{n}^{n'}\delta_{k}^{k'}
 \ +\
 \frac{2\hat{W}}{\hat{\eta}\sqrt{2\pi \hat{c}}}\
 \sum_{i}^{N_s}\
 \hat{M}_{nm}^{n'm'}(\hat{\bf r}_i)\ \hat{\varepsilon}_i.
 \label{hamiltonian_projected_2}
\end{equation*}
To study the limit of vanishing magnetic field 
for a given fixed disorder potential [Eq.~(\ref{potential_1})], 
it is equivalent to consider the limit $\hat{\eta}\to 0$ 
in the dimensionless Hamiltonian. Therefore, in some respect it corresponds to
a white noise disorder model which does not discern between 
$B=\mathrm{const,}\ \eta\to 0$ and $\eta=\mathrm{const,}\ B\to 0$. 
However, the normalization of the disorder potential may be chosen differently 
from the $\delta$-property often used in studies of the high magnetic field limit.

For the disordered system projected onto the lowest two Landau levels we 
find the following matrix elements for a single scatterer. 
Inside the lowest Landau level we get
with the correlation parameter $\hat{\sigma}^2=\hat{\eta}^2+1=(\eta/l_B)^2+1$ 
\begin{eqnarray}
\lefteqn{\hat{M}_{0m}^{0m'} \ =\ 
 \frac{\sqrt{\pi}(\hat{\sigma}^2-1)}{\hat{L}_x \hat{\sigma}}\
 e^{-\frac{\hat{\sigma}^4-1}{4\hat{\sigma}^2}
     (\frac{2\pi}{\hat{L}_x^2})^2\,(m-m')^2}\ \times}  \\
&&
 e^{-\frac{(2\pi)^2}{2\hat{\sigma}^2\hat{L}_x^2}
     \left((m-Y_0)^2+(m'-Y_0)^2\right) 
     \,+\,
     i\,(\frac{2\pi}{\hat{L}_x^2})^2 \,(m-m') X_0 }
 \,.
 \nonumber
\end{eqnarray}
All higher intra and inter Landau level interaction matrix elements can be 
expressed as a polynomial in $m-Y_i$, $m'-Y_i$ and $m-m'$ times the
intra-lowest Landau level elements. 
For the inter Landau level matrix elements we get
\begin{eqnarray}
\lefteqn{\hat{M}_{1m}^{0m'} \ =\ 
 \left(\hat{M}_{0m'}^{1m}\right)^* \ =\ } \\
&&
 -\frac{\sqrt{2}}{\hat{\sigma}^2}\frac{2\pi}{\hat{L}_x}\
 \Big( (m-Y_i) +\frac{\hat{\sigma}^2-1}{2}\,(m-m') \Big)\
 \hat{M}_{0m}^{0m'},
 \nonumber
\end{eqnarray}
and inside the second Landau level
\begin{eqnarray}
\lefteqn{\hat{M}_{1m}^{1m'} \ =\ 
 \frac{2}{\hat{\sigma}^4}\
 \hat{M}_{0m}^{0m'} \ 
 \Bigg\{\ \frac{\hat{\sigma}^2(\hat{\sigma}^2-1)}{2} \ +\ } \\
&&
 \frac{(2\pi)^2}{\hat{L}_x^2}\
 \Big((m-Y_i)(m'-Y_i)-\frac{\hat{\sigma}^4-1}{4}\,(m-m')^2 \Big) 
 \ \Bigg\}
 \,.
 \nonumber 
\end{eqnarray}
Here, $Y_i=\hat{y}_i\,\hat{L}_x/2\pi$ and $X_i=\hat{x}_i\,\hat{L}_x/2\pi$ 
are the spatial coordinates of the particular scatterer. 
We write the integer $m$ instead of the momenta $\hat{k}=2\pi\,m/\hat{L}_x$
and restrict them to the range $m\in[0,N_{\hat{k}}-1]\subset Z_1$, which
renders the sample area finite. 
For a square sample we have $\hat{L}_y=\hat{L}_x$ 
and $\hat{L}_x^2=2\pi\,N_{\hat{k}}$, 
thus $Y_i,X_i\in[0,N_{\hat{k}}-1]\subset R_1$. The concentration 
of the scatterers has been chosen in the range 
$c={N_s/(L_x L_y)} = 10\ldots40$, 
high enough to reach the high concentration limit in the density of states.

For the Hamiltonian in the restricted $n,m$-base, the energy eigenvalues
were now computed numerically by exact diagonalization for
$N_{\hat{k}}=48^2/2$ and various correlation length $\hat{\eta}$ 
and disorder strength $\hat{W}$.
The positions of the extended states have been extracted by means of 
the level statistics method.\cite{Sea93,BSZK96} 

\section{Results for the two-level system}
A quantum Hall model projected onto the lowest Landau level only (one
band model) still exhibits localization and scaling near the critical states 
at the band center\cite{HK90}. This behavior was observed also in the presence
of spatially correlated disorder potentials.\cite{Huc95} Of course, the one
band model neither shows levitation of the critical states nor any movement 
of the density of states (DOS) peak position when the magnetic field or the 
disorder strength is varied. Therefore, we consider in the following the case 
of a system projected onto the lowest two Landau levels.

\begin{figure}
\centerline{\includegraphics[width=8.cm]{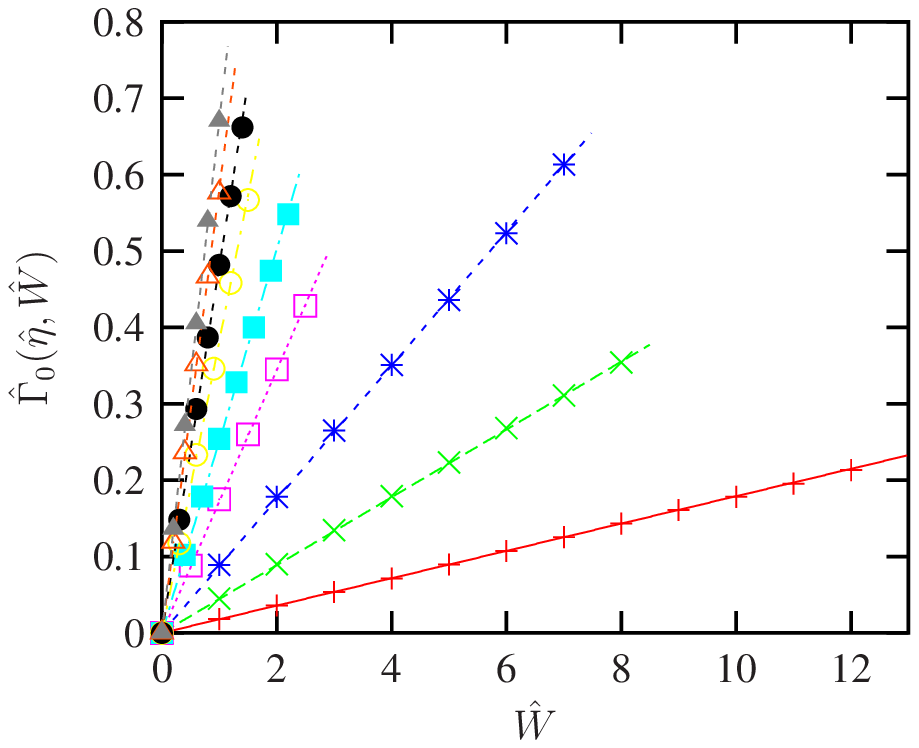}}
 \caption[]{(Color online) The linear broadening of the lowest Landau
  band's width $\hat{\Gamma}_0(\hat{\eta})$ versus the strength
  $\hat{W}$ of the disorder potential. The potential correlation lengths 
  $\hat{\eta}$ are 0.02 ($+$), 0.05 ($\times$), 0.1 ({\large$\ast$}), 
  0.2 ({\small$\Box$}), 0.3 ($\rule{1.7mm}{1.7mm}$), 0.5 ({\large$\circ$}), 
  0.7 ({\large$\bullet$}), 1.0  ({\footnotesize$\triangle$}), and 1.5 
  ($\blacktriangle$), respectively.}
\label{Gamma0}
\end{figure}

\begin{figure}[b]
\centerline{\includegraphics[width=8.cm]{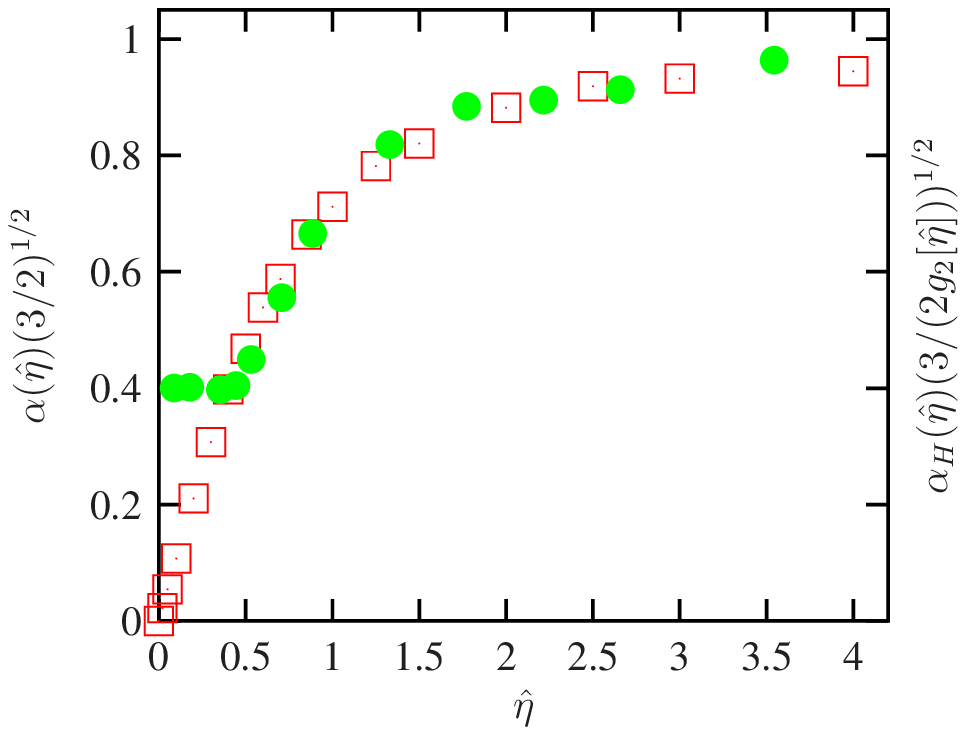}}
 \caption[]{(Color online) The coefficient $\alpha(\hat\eta)$
  ({\small$\Box$}) from
  the relation $\hat{\Gamma}_0 =\alpha(\hat\eta)\,\hat{W}$,
  which collapses the linear broadening of the lowest disordered
  Landau band onto a single curve (see Fig.~\ref{a-collapse}),
  is shown as a function of the reduced correlation length
  $\hat{\eta}$. The result from a Harper model ({\large $\bullet$}) 
  ($B=1/8, \eta=0.1\ldots4.0$) deviate only for $\hat{\eta}\lesssim 0.5$.
  To account for the lattice effects in the Harper data, actually 
  $(3/2)^{1/2}\alpha(\hat\eta)$ (left y-axis) and
  $(3/(2g_2[\hat\eta]))^{1/2}\alpha_\mathrm{H}^{}(\hat\eta)$ 
  (right y-axis) are 
  shown (for $g_2[\hat{\eta}]$ see Ref.~\onlinecite{KS03}), which
  relate the width $\hat{\Gamma}_0$ of the lowest Landau band 
  to the width of the random potential distribution
  for the continuum and the Harper model, respectively.
  }
 \label{plot-alpha}
\end{figure}

\begin{figure}[t]
\centerline{\includegraphics[width=7.5cm]{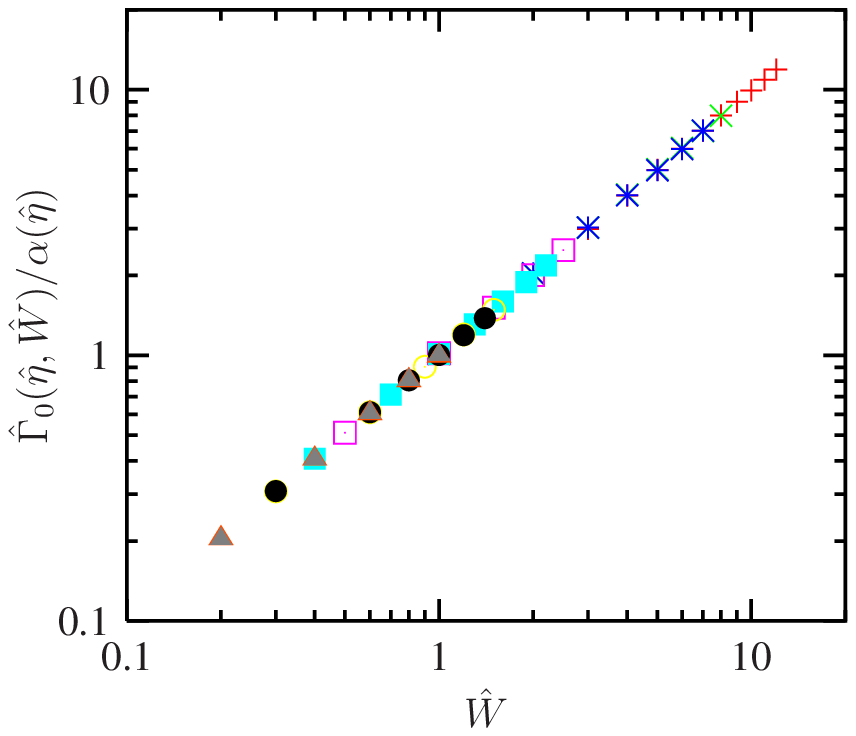}}
 \caption[]{(Color online) The broadening of the lowest Landau band's 
  half width $\hat{\Gamma}_0(\hat{\eta},W)=\alpha(\hat{\eta})\,\hat{W}$ 
  versus disorder strength $\hat{W}$. All data for various correlation
  lengths $\hat{\eta}$ fall onto a straight line.
  }
 \label{a-collapse}
\end{figure}

With increasing strength $\hat{W}$ of the disorder potential the Landau
levels broaden into bands. Their width increases, with good accuracy,
linearly with the disorder strength $\hat{W}$ for all correlation
lengths $\hat\eta$ considered. The ratio of the broadening of the first
and the second band increases with increasing correlation length, 
starting at nearly equal broadening for small $\hat\eta$. For a fixed 
second moment of the disorder potential the width of the bands
increases with $\hat\eta$.

\begin{figure}[b]
\centerline{\includegraphics[width=7.5cm]{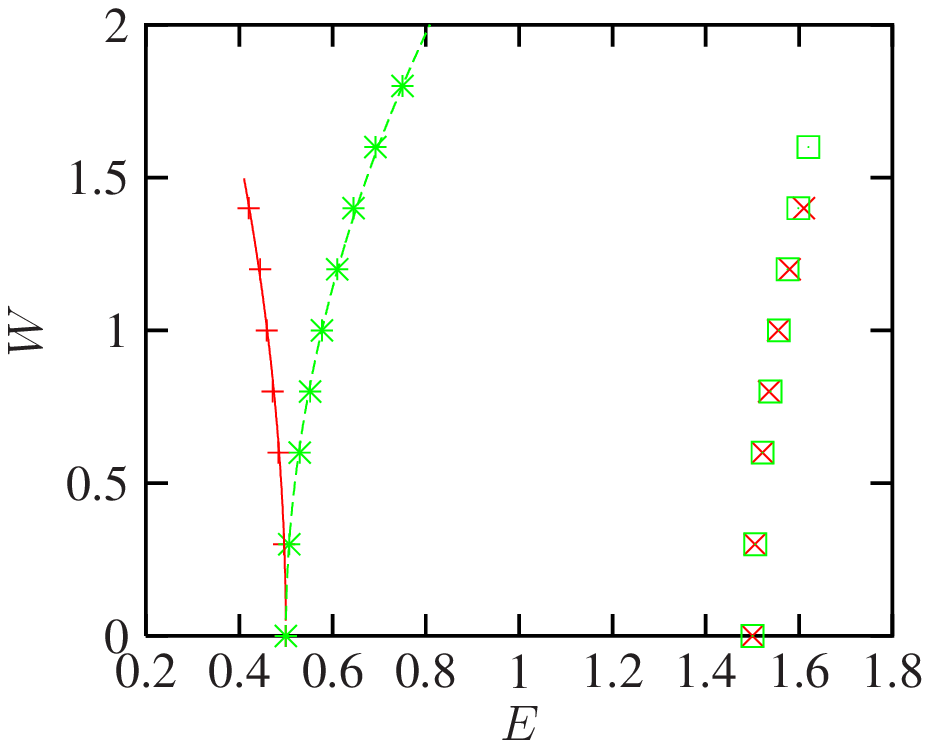}}
 \caption{(Color online) The energy $E$ of the position of the Landau 
  band peaks ($+$, $\times$) and of the extended states ({\large$\ast$}, 
  $\Box$)
   as a function of disorder strength $W$ for the projected continuum 
   model incorporating the lowest two unperturbed Landau levels.
   The lines are quadratic fits $\delta E\propto W^2$.}
 \label{plot_fork}
\end{figure}

In Fig.~\ref{Gamma0} the full broadening $\hat{\Gamma}_0$ at half maximum 
is shown for the 
lowest Landau band. Because of its uniform behavior, we can use the 
width of the lowest Landau band as a measure for the effective disorder 
strength to compare various correlation lengths $\hat\eta$.\cite{KS03} 
Therefore we define a function $\alpha(\hat\eta)$ which allows to 
collapse all $\hat{\Gamma}_0(\hat{W})$ for
different $\hat\eta$ onto a single curve $\hat{\Gamma}_0 = 
\alpha(\hat\eta)\,\hat{W}$.
The function $\alpha(\hat\eta)$ is plotted in Fig.~\ref{plot-alpha}
together with the numerical data obtained for the disordered Harper 
model for comparison. The continuum and the lattice model
differ only for $\hat{\eta}\lesssim 0.5$. 
Note that $\alpha(\hat\eta)$ for large $\hat\eta$ approaches $\sqrt{2/3}$
as has been previously reported for the Harper model.\cite{KS03}
Our data show a strictly linear dependence $\alpha(\hat\eta)\propto\hat\eta$
for small $\hat\eta$.
The resulting data collapse is shown in Fig.~\ref{a-collapse}.

\begin{figure}[t]
\centerline{\includegraphics[width=7.5cm]{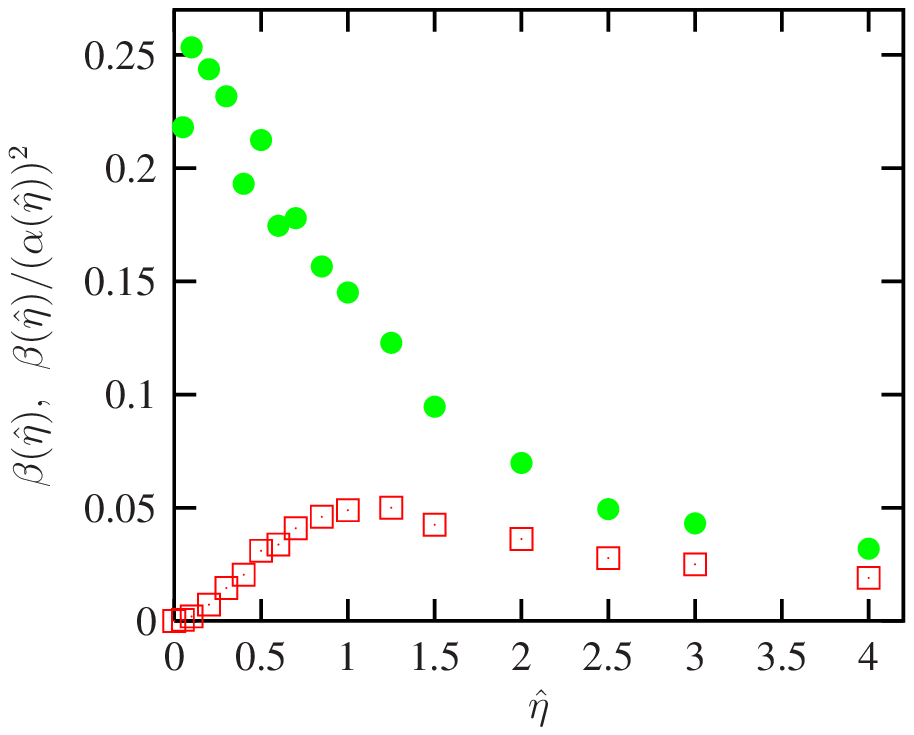}}
 \caption[]{(Color online)
   The coefficient $\beta(\hat\eta)$ ($\Box$) 
   which describes the energy shift of the DOS peak of the lowest Landau band, 
   $\delta\hat{E}_{\rm DOS} = -\beta(\hat{\eta})\,{\hat W}^2$, 
   and the coefficient $\beta(\hat{\eta})/(\alpha(\hat{\eta}))^2$
   ({\large $\bullet$}) 
   relating $\delta\hat{E}_{\rm DOS}$ and $\hat{\Gamma}_0^2$.
   The corresponding data collapse onto a single curve is shown in
   Fig.~\ref{plot_beta_coll} for various correlation lengths $\hat{\eta}$.
   }
 \label{plot_beta}
\end{figure}

\begin{figure}[b]
\centerline{\includegraphics[width=7.5cm]{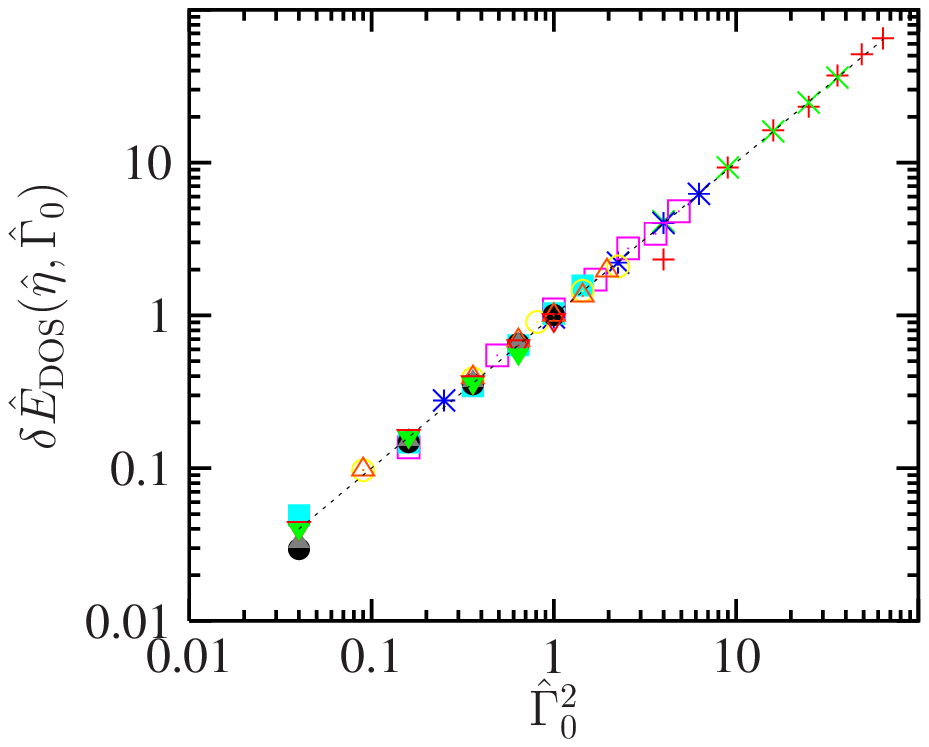}}
 \caption[]{(Color online)
   The energetic shift of the density of states' peak position versus
   the square of the lowest Landau level broadening $\hat{\Gamma}_0$.
   The data for various correlation lengths $\hat{\eta}$ fall onto a 
   straight line, $\delta\hat{E}_{\rm DOS}=
   \beta(\hat{\eta})/(\alpha(\hat{\eta}))^2 \hat{\Gamma}_0^2$
   }
 \label{plot_beta_coll}
\end{figure}

The typical effect of increasing disorder on the the peak positions of 
the two Landau bands as well as on the positions of the extended 
states is shown in Fig.~\ref{plot_fork}. The DOS peak of the lowest 
Landau band moves down in energy, away from the second band, 
which moves to higher energies with increasing disorder strength. 
The extended states behave differently. 
While those of the second band are essentially following the DOS
peak, the lowest band's extended states clearly float up in energy, 
which can be traced approximately across 20\% of the Landau gap.
This behavior is qualitatively independent of the correlation length
$\hat\eta$. 

As we cannot expect the results for the second Landau band to be physically 
meaningful within a model projected only onto the two LLLs, 
we concentrate in the following on the properties of the lowest Landau band.
Throughout the investigated range $\hat\eta= 0.05 \ldots 4.0$ of correlation lengths
the shift of the peak position of the density of states to lower energy 
is with reasonable accuracy $\propto\hat{W}^2$ 
and of considerable magnitude, comparable with the
magnitude of the extended states' shift absolute in energy.
We therefore define a function $\beta(\hat{\eta})$ which collapses all 
$\delta\hat{E}_{\rm DOS}(\hat W)$ for different $\hat{\eta}$ onto a single
curve $\delta\hat{E}_{\rm DOS}=-\beta(\hat{\eta})\,{\hat W}^2$.
The function $\beta(\hat{\eta})$ is shown in Fig.~\ref{plot_beta}, with
the collapse of the data shown in Fig.~\ref{plot_beta_coll}.
For a fixed second moment of the disorder potential the DOS peak shift has 
a maximum around $\hat\eta\approx 1$ 
and decays quadratically for $\hat\eta\to 0$.
If the width of the lowest Landau band $\hat{\Gamma}_0$ is used as an effective
disorder strength, the coefficient 
$\beta(\hat{\eta})/(\alpha(\hat{\eta}))^2$ of the shift of the density of states
does not diverge for small $\hat{\eta}$ but
rather behaves linearly, approaching a finite value for $\hat{\eta}=0$.

\begin{figure}[t]
\centerline{\includegraphics[width=7.75cm]{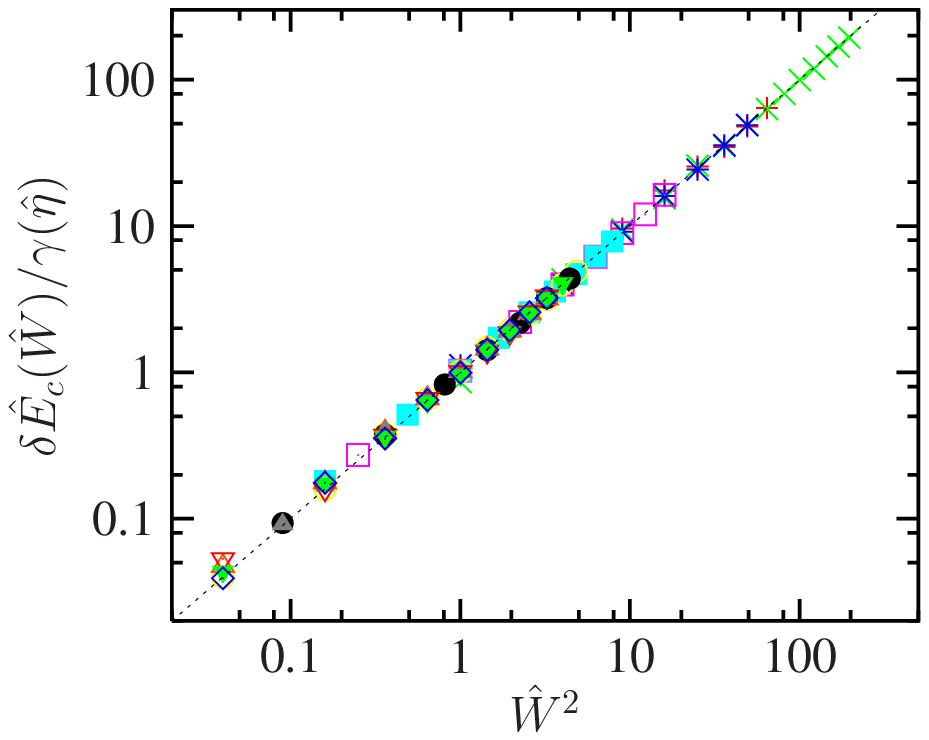}}
 \caption[]{(Color online) The levitation of the lowest Landau bands' 
   extended states with increasing disorder strength
   $\delta\hat{E}_c=\gamma(\hat\eta)\,\hat{W}^2$,
   collapsed onto a single curve by the function $\gamma(\hat\eta)$,
   is shown for several correlation length $\hat{\eta}$.
   }
 \label{plot_gamma_coll}
\end{figure}

\begin{figure}[b]
\centerline{\includegraphics[width=7.5cm]{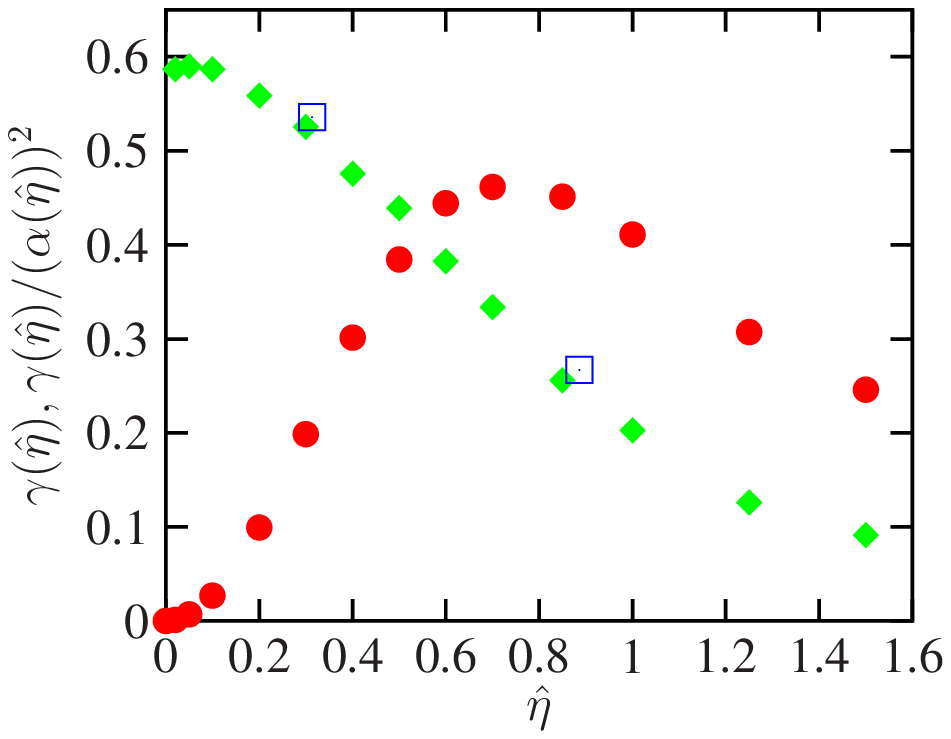}}
 \caption[]{(Color online) 
   The coefficients $\gamma(\hat\eta)$ ({\large $\bullet$} multiplied 
   by a factor 6), which collapses the data onto a straight line, 
   and $\gamma(\hat\eta)/(\alpha(\hat\eta))^2$
   (\rotatebox{45}{$\rule{1.5mm}{1.5mm}$})
   are shown as a function of the correlation length $\hat{\eta}$. 
   The data points ($\Box$) are the corresponding 
   results from the projected Harper model discussed below.}
 \label{plot_gamma}
\end{figure}

The absolute levitation of the lowest Landau band's extended states with 
increasing disorder strength measured via $\hat{\Gamma}_0$ was calculated
for several $\hat{\eta}$. The shift of the extended states' positions in 
energy is $\propto\hat{\Gamma}_0^2$ and thus proportional to the disorder 
potentials second moment $\hat{W}^2$ for all $\hat\eta$. 
Introducing a scaling function $\gamma(\hat\eta)$, we can again collapse 
the data for all $\hat\eta$ onto a single curve 
$\delta\hat{E}_c = \gamma(\hat\eta)\,{\hat W}^2$ 
as shown in Fig.~\ref{plot_gamma_coll}.
Apart from the numerically least resolved points at large disorder strength, 
for each $\hat{\eta}$ the data fit very well onto a single quadratic curve. 
We have to remark that because of the stronger broadening of the Landau bands 
for larger $\hat{\eta}$, larger shifts on this common curve can be resolved for 
smaller $\hat{\eta}$.

The functions $\gamma(\hat{\eta})$ and $\gamma(\hat{\eta})/(\alpha(\hat{\eta}))^2$ 
relating the levitation $\delta\hat{E}_c$ to $\hat{W}^2$ and 
$\hat{\Gamma}_0^2$, respectively, are shown in Fig.~\ref{plot_gamma}. 
$\gamma(\hat{\eta})$ decays quadratically for $\hat{\eta}\to 0$, shows a maximum 
around $\hat{\eta}\approx 0.7$, and decreases monotonically for larger $\hat{\eta}$.
Therefore the levitation for a given broadening $\hat{\Gamma}_0$ of the lowest 
Landau band, $\gamma(\hat{\eta})/(\alpha(\hat{\eta}))^2$, is finite at 
$\hat{\eta}=0$ and decreases with increasing correlation length $\hat{\eta}$.

\section{Comparison with projected Harper model}
\begin{figure}
\centerline{\includegraphics[width=8.cm]{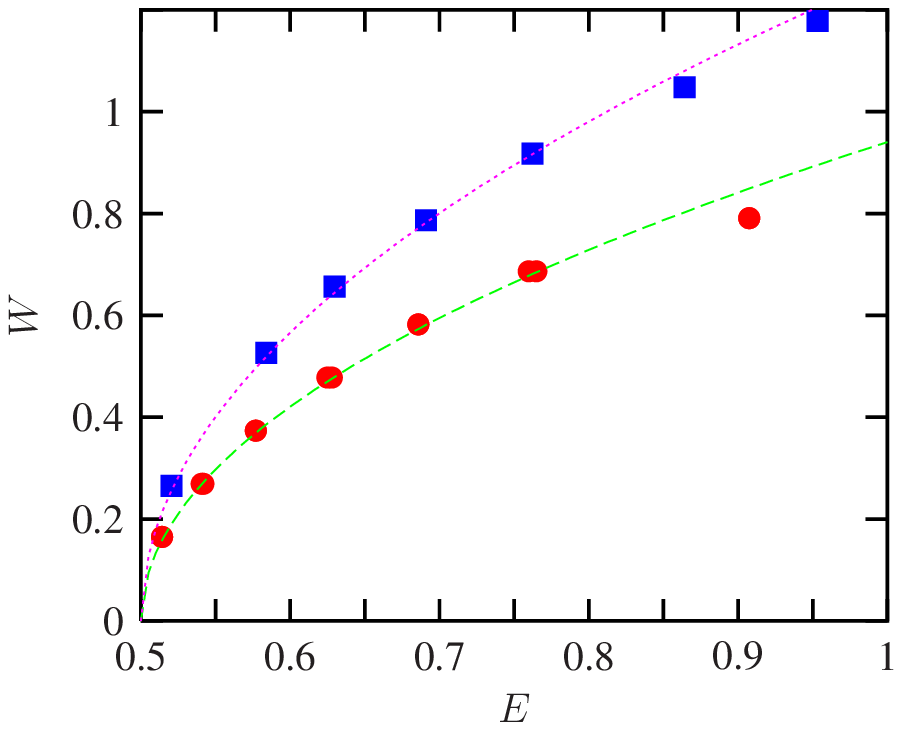}}
 \caption[]{(Color online)
   The floating of the energetical position of the lowest
   extended states with increasing disorder strength is shown for the 
   two-band projected disordered Harper model. 
   The parameters are $B=\phi_0/(64a^2),\ \eta=1.0a$ ({\large $\bullet$}),
   and $B=\phi_0/(32a^2),\ \eta=2.0a$ ($\rule{1.7mm}{1.7mm}$)
   which correspond to $\hat\eta=0.31$ and $\hat\eta=0.89$, respectively.
   The lines are quadratic fits $\delta\hat{E}\propto\hat{\Gamma}_0^2$.
 }
 \label{plot_harper}
\end{figure}

Before we start the discussion about the important limit $B\to 0$, we would
like to compare our results obtained for the continuum model with those 
from a projected disordered Harper model.\cite{KS03} 
The latter shows floating across almost half the Landau gap.
In Fig.~\ref{plot_harper} the floating of the energetical position of the
lowest extended states are shown in a similar way as in Fig.~\ref{plot_fork}
for two magnetic flux densities $B=\phi_0/(64a^2), \eta=1$, and 
$B=\phi_0/(32a^2), \eta=2$, respectively.
For ease of comparison, the energy scale of the Harper model has been shifted, 
$E=E_{\mathrm H}+0.5-E^0_{\mathrm H}$, 
where $E^0_{\mathrm H}$ is the energy of the lowest extended states in the
unperturbed Harper model, and $\phi_0=h/e$ denotes the flux quantum and 
$a$ the lattice constant.

As in the projected continuum model, the extended states float up in
energy $\delta\hat{E}_c=\gamma(\hat\eta)\,\hat{W}^2=
\gamma(\hat\eta)/(\alpha_\mathrm{H})^2\,\hat{\Gamma}^2$. The perfect
agreement of both models can be seen from  
Fig.~\ref{plot_gamma} where the two data points from the Harper model 
($\Box$) are shown to fall onto the respective curve
for the continuum model. The addition of further disorder broadened 
sub-bands reduce the floating of the extended states \cite{KS03} and 
produce deviations from the $\delta\hat{E}_c\propto\hat{W}^2$
behavior which is most evident for the full disordered Harper system.
We expect similar effects to occur for the continuum model 
when more than the two lowest Landau bands are taken into account.

\section{Discussion}
The projection onto the lowest two Landau levels is the most simple system
which shows levitation of at least the lowest Landau band's extended states.
Previous work \cite{KS03}
has suggested that the main contribution to the levitation of the 
lowest level results from the coupling to the second band and should have
a quadratic dependence on disorder strength. 
The advantages of the continuum model discussed in the present paper are 
\textit{(i)} that we have no interfering intrinsic lattice effects as in 
the projected Harper models, and 
\textit{(ii)} that we can absorb the magnetic field dependence into the units
of energy and length, such that the only parameters of the model are the
disorder potential strength and its correlation length measured in units of 
the magnetic length. 
This reduction to a dimensionless $\hat\eta$-parameterized model is exact for
an infinite system.
By calculating the properties of the dimensionless model for decreasing 
$\hat\eta$ we can learn about the behavior of the original system 
for fixed disorder potential and decreasing magnetic field.

Let us now consider the limit of vanishing magnetic field. Given a fixed disorder
potential, as it occurs in most experiments on samples without an additional 
backgate, this means the limit $\hat{\eta}\to 0$.
For the width $\hat{\Gamma}_0=\alpha(\hat{\eta})\,\hat{W}$ 
of the disorder broadened lowest Landau band, we find from the 
function $\alpha(\hat\eta)$ at small $\hat\eta$ (see figure \ref{plot-alpha})
a linear relation $\hat{\Gamma}_0\sim\hat{\eta}\,\hat{W}$. 
In natural units this reads $\Gamma_0\propto W\sqrt{B}$, 
which means that the disorder induced broadening of the Landau bands 
decreases for a given physical disorder potential 
with vanishing magnetic field $B$, but slower than the Landau gap.
Hence, the overlap of adjacent Landau bands increases for $B\to 0$.
This is qualitatively in agreement with previous work.\cite{AU74,Weg83}
The levitation of the extended states 
$\delta\hat{E}_c=\gamma(\hat\eta)\,\hat{W}^2$ 
for small $B$ is ruled by the behavior
of the function $\gamma(\hat\eta)$ near $\hat\eta\approx 0$. 
Since $\gamma(\hat\eta)\sim\hat\eta^2$ for small $\hat\eta$, 
the coefficient
$\gamma(\hat\eta)/(\alpha(\hat\eta))^2={\rm const}+o(\hat\eta)$ 
for the levitation of the extended states 
$\delta\hat{E}_c\propto \hat{\Gamma}_0^2$
starts as a constant (see Fig.~\ref{plot_gamma}). 
Hence, the shift of the extended states' energy is confined 
by the broadening of the Landau levels DOS, 
$\delta\hat{E}_c\approx {\rm const} \cdot \hat{\Gamma}_0^2$.

In experiments, where the second moment of the disorder potential is fixed 
in natural units, the relative broadening $\hat{\Gamma}_0$ diverges 
for small magnetic field $\hat{\Gamma}_0\sim W/\sqrt{B}$ which leads to an
infinite shift of the relative Energy $\delta\hat{E}_c\sim W^2/B$ of the 
lowest Landau band's extended states. 
As a consequence the critical filling factor, i.e., 
the number of electrons divided by the number of flux quanta penetrating
the area of the system where the Fermi energy coincides with the energy of
the extended states, diverges as well for $B\to 0$.
However, if we consider the energy shift $\delta E_c$ of these extended 
states in natural units, the $B$-dependence cancels for small $B$ 
and we find with $\hat{\eta}^2\propto \eta^2 B$ a {\em finite} levitation 
$\delta E_c = \gamma(\hat\eta)/(\hbar\omega_c)\,W^2 %
 \sim \mathrm{const}\cdot W^2$ 
that only depends on the properties of the disorder potential. 
This result clearly conflicts with the levitation scenario proposed by 
Khmelnitskii \cite{Khm84} and Laughlin \cite{Lau84} which predicts a
divergent energy of the extended states for vanishing magnetic field.
In order to obtain such a divergency from our model,
the function $\gamma(\hat\eta)/(\alpha(\hat\eta))^2$ 
would have to diverge like $o(\hat{\eta}^{-2})$, 
hence $\gamma(\hat\eta)\sim\mathrm{const}$, for small $\hat\eta$.
Here, we assumed the scattering time $\tau$ in the 
Khmelnitskii-Laughlin expression for the energy level float up
to be independent of $B$. There are, however, attempts to describe
the classical low field magnetoresistance in terms of a $B$-dependent
$\tau$.\cite{BMHH95,DDJ01,DDJ02}

Interpreting these results we have to consider several issues.
First of all, in the present work we included only the lowest two Landau bands.
Though this is certainly reasonable for small broadening of the Landau bands
because the strongest influence on the first Landau band's extended states
comes from the inter-band matrix elements that couple the first to the 
second Landau level.
Yet, it is not clear that the influence of the higher levels remains small for 
stronger overlapping bands.
From projected disordered Harper models including the lowest two and the
lowest three Harper bands,\cite{KS03} we have some evidence that the addition of
the third band reduces the levitation of the first band's extended states. 
The dependence on the disorder strength remains quadratic in this case.
On the other hand for strong magnetic fields, the levitation of the lowest 
band's extended states in full disordered Harper models has shown to be even 
slower and not quadratic in the disorder strength anymore. 
However, for those models the levitation of the extended states could be traced
at least across the Landau gap, which is far beyond the resolvable range in the
continuum model discussed in the present paper. 
Nevertheless, it remains to be seen, whether the finite levitation result
obtained here will persist when taking into account more than the 
lowest two Landau bands.

Second, the possible influence of the finite system size. The elimination 
of the explicite $B$ dependence of the system by the reduction to the 
$\hat\eta$-parameterized
dimensionless model is only exact for an infinite system.
For the numerical simulation we have to limit the Hilbert space 
which is done by restricting the number of Landau levels $n$ and, 
particularly, the number of momenta $k$ in the representation of the 
projected Hamiltonian [Eq.~(\ref{hamiltonian_projected_1})]. 
The limitation of the $k$ effectively renders the area of the sample,
$F\approx  k_{\rm max} L_x l_B^2$, finite and the dimensionless model
an approximation. However, the finite size of the sample is expected to
affect only large $\hat\eta$, or large $\eta$ for a given $B$,
not the opposite limit $\hat\eta\to 0$ which we are discussing here.
Note that the effective (linear) size of the system scales with $l_B$. 
The other approximation we made with respect to the system size is the
restriction of the spatial integration domain over the Gaussian scattering 
potentials in the calculation of the matrix elements. 
We approximated the finite integration domain in the numeric simulation
by indefinite Gaussian integrals. 
Again this should not affect the limit of short correlation lengths.
In a finite sample the simulation results will depend on the concentration
$c$ of the scatterers in the random potential [Eq.~(\ref{potential_1})]
albeit the local statistical properties of the potential do not.
In our calculations we have chosen increasing $c$ for smaller $\hat\eta$,
high enough to reach the high-concentration limit in the density of states.
Because of the limited available computer power it was not feasible to
directly study the anticipated corresponding convergence in the behavior of
the critical states with increasing scatterer concentration.
This should be addressed in future work.

Third, the behavior has been extrapolated from relatively small but non-zero 
$\hat{\eta}$. Although there is no indication for a discontinuous behavior
when $\hat\eta\to 0$, there is no guarantee that we can extrapolate our data
that far. However, the range of computed $\hat{\eta}$ seems to be comparable
with experiments. For instance, combining atomic force microscopy with
selective etching, the image of the topology of interfaces in AlGaAs/GaAs 
quantum well structures was shown to exhibit smoothly varying structures 
with correlation length in the range 
4$\cdot 10^{-8}$\,m to 2$\cdot 10^{-7}$\,m.\cite{Gea03} 
For the latter value, this leads to approximately $0.1\le\hat{\eta}\le 10$ 
when the magnetic flux density is varied between $10^{-4}$ and 1 Tesla. 

Our findings for the disorder broadening of the lowest Landau level
are in qualitative agreement with results obtained previously by 
Ando and Uemura.\cite{AU74} The basic difference to our work is the 
incompatible definition of the disorder potential: Our $V(r)$ is normalized
to a given second moment, whereas Ando and Uemura normalize their disorder 
potential to a fixed integral $\int V(r)\ {d^2r}$ ($\delta$-property).
This leads to an additional factor $\hat\eta$ in the potential strength, 
$\hat{W}_\mathrm{AU}\propto \hat\eta\,\hat{W}$.
In the limit of small correlation length $\hat{\eta}\to 0$,
we find for the broadening of the lowest Landau level a linear 
$\hat{\eta}$-dependence, $\Gamma_0\sim \hat{\eta}\,W$, 
which directly translates into Ando and Uemura's independence 
 on the correlation length, $\Gamma_0\sim W_\mathrm{AU}$.
In the opposite limit of large $\hat{\eta}$, we find that the broadening of the
DOS just corresponds to the second moment of the disorder potential, again
perfectly in agreement with their results.
For the behavior of the critical states in the limit $B\to 0$, not considered 
by Ando and Uemura, the alternative normalization of the potential 
strength may lead to significantly different results. 
In our case, we consider it physically correct to extract the $B$-dependence 
of the energy of the critical states from the $\hat{\eta}$-dependence of the 
model with a fixed second moment $W^2/3$ of the disorder potential in 
natural units. 
If we used instead the $\hat{\eta}$-dependence for a preserved $\delta$-property
of the potential, i.e., $\int V(r)\ {d^2r}=\mathrm{const}$ 
as for the potential used by Ando and Uemura,\cite{AU74} we would obtain a singular 
dependence $\delta E_c\propto W^2_\mathrm{AU}/B$ for $B\to 0$ which corresponds 
better to the levitation scenario proposed by Khmelnitskii and Laughlin.

\section{Summary}
We investigated a disordered two-dimensional continuum model for the  
integer quantum Hall effect projected onto the lowest two Landau levels. 
The chosen dimensionless representation enabled us to make
statements about the energetic shift of both, the magnetic field and 
correlations length dependence of the density of states peak and of the 
extended states. The results were obtained from a numerical investigation 
of the model, parameterized by the strength $\hat{W}$ and the correlation 
length $\hat{\eta}$ of an effective disorder potential only. The magnetic 
field $B$ dependence was entirely absorbed into the units of length and 
energy.

For the two-level continuum model we found qualitatively the same behavior
as has been observed for disordered Harper models projected onto the lowest
two sub-bands.\cite{KS03} With increasing disorder strength $\hat{W}$ the 
Landau bands broaden linearly. We clearly observed floating up in energy 
of the lowest Landau band's extended states while the peak of the density 
of states moves downwards. Both effects are proportional to the square of 
the disorder strength.

In the limit $B\to 0$, which is difficult to access in numerical studies on 
lattice models, we find for the projected two-band continuum model that the 
energy of the extended states only floats up to a finite value that depends 
on disorder.
As this result is at variance with the diverging floating scenario suggested
20 years ago\cite{Khm84,Lau84}, we discuss a few limitations of our work that 
may be responsible for the discrepancy. On the other hand, our result is in 
accord with some experimental work. Still, the addition of higher Landau
levels and the investigation of the relevant matrix elements are clearly
needed to settle this issue and to gain a deeper insight into the microscopic 
origin of the levitation of the extended states and the corresponding quantum 
Hall to insulator transition. We feel that to follow up our approach within 
a projected continuum model will render this objective possible. 

\section{Acknowledgments}
We thank Bodo Huckestein for valuable discussions.
This research was supported by the DFG-Schwerpunktprogramm
``Quantum-Hall-Systeme'' Grant No. SCHW795/1-1.

\bibliographystyle{apsrev}

\end{document}